\documentclass[10pt,journal]{IEEEtran}
\usepackage{amsmath,amsfonts}

\usepackage{amsthm,amssymb}

\usepackage{bm}
\usepackage{algorithm}
\usepackage{algpseudocode}
\usepackage{array}
\usepackage{subfigure}
\usepackage{textcomp}
\usepackage{stfloats}
\usepackage{url}
\usepackage{verbatim}
\usepackage{graphicx}
\usepackage{cite}
\usepackage{color}

\begin{document}

\title{\huge Communication-Learning Co-Design for Differentially Private Over-the-Air Federated Distillation}

\author{\normalsize Zihao Hu, \textit{Student Member, IEEE}, Jia Yan, \textit{Member, IEEE}, Ying-Jun Angela Zhang, \textit{Fellow, IEEE}\thanks{Z. Hu and Y.-J. A. Zhang are with the Department of Information Engineering, The Chinese University of Hong Kong, Hong Kong SAR. (e-mail: \{hz021; yjzhang\}@ie.cuhk.edu.hk). J. Yan is with the Intelligent Transportation Thrust, The Hong Kong University of Science and Technology (Guangzhou), Guangzhou, China. (e-mail: jasonjiayan@hkust-gz.edu.cn).}}
\maketitle
\begin{abstract}
    The ever-growing learning model size nowadays challenges the communication efficiency and privacy preservation of the traditional federated learning (FL). In this paper, we propose a novel differentially private (DP) over-the-air federated distillation (FD) framework, where wireless devices (WDs) periodically share noise-perturbed model outputs with the parameter server by harnessing the superposition property of multi-access channels. Accordingly, over-the-air FD enables the shared responsibility of the DP preservation on the low-dimensional disclosed signals among WDs. We study the communication-learning co-design problem in differentially private over-the-air FD, aiming to maximize the learning convergence rate while meeting the transmit power and DP requirements of WDs. The main challenge is rooted in the intractable learning and privacy analysis in over-the-air FD, together with the strong coupling among the decision variables spanning two timescales. To tackle this problem, we first derive the analytical learning convergence rate and privacy losses of WDs, based on which the optimal transceiver design per FD round and long-term training rounds decision are obtained in the closed forms. Numerical results demonstrate that the proposed differentially private over-the-air FD approach achieves a better learning-privacy trade-off with largely-reduced communication overhead than the conventional FL benchmarks.
\end{abstract}
\section{Introduction}
Federated learning (FL) has been recognized as a promising distributed edge learning paradigm, where wireless devices (WDs) periodically share their model updates with the parameter server (PS) while keeping datasets locally \cite{FL}. To improve the communication efficiency of FL at the wireless edge, over-the-air computation is leveraged by harnessing the inherent superposition property of multi-access channels. Through simultaneous transmission of analog-modulated local model updates via the same time-frequency resource, over-the-air FL is scalable to support a large number of WDs \cite{OTA}.\par
On the other hand, although FL avoids raw training data sharing, researches on the model inversion attack show that the transmitted model updates can leak private information of local datasets. Motivated by this, differential privacy (DP) has been investigated to quantify and control the privacy leakage of WDs in the FL system. To preserve DP of WDs in FL, the authors in \cite{DP} applied the Gaussian mechanism by imposing the well-designed Gaussian artificial noise on the disclosed local model updates. Compared with the orthogonal model aggregation in conventional FL, over-the-air FL facilitates shared responsibility among WDs for the privacy preservation since the DP noise from each WD is naturally superposed \cite{DP-FL}. The optimal device sampling strategy is investigated in \cite{D-FL} to attain a better trade-off between the learning performance and privacy preservation in over-the-air FL. Typically, the amount of DP noise is proportional to the model dimension.\par
With the emergence of AI models with the increasing number of parameters, e.g., large language models and deep residual networks, the existing over-the-air FL is far away from communication efficiency and DP preservation. To this end, federated distillation (FD) with the synergy of the knowledge distillation and FL becomes a promising solution under large size training models. By leveraging the model output as the medium for knowledge exchanging, the dimension of the transmitted signal of each WD is only related to the number of sample classes in the local dataset, regardless of the model size. Over-the-air FD is studied in \cite{OTA-FD} by optimizing the transceiver design to maximize the FD learning performance. Furthermore, with the largely-reduced signal dimension, FD demonstrates lower privacy leakage compared to the conventional FedAvg framework under white-box privacy attacks \cite{S-FD}.\par
In this regard, over-the-air FD has great potential to further enhance the communication efficiency and privacy for collaborative edge learning. How to achieve a better learning-privacy trade-off is crucial for over-the-air FD. This calls for the joint optimization of communication and learning strategies, which poses the following threefold challenges. First, the DP noise, wireless channel fading and communication noise perturb the local model outputs instead of model parameters, rendering the existing FL analysis inapplicable. Second, model outputs after the softmax layer form a probability simplex for each sample class, which exacerbates the difficulty of the DP analysis. Lastly, the transceiver design per FD round is strongly coupled with the long-term total number of training rounds design.\par
To address the above challenges, this paper makes the first attempt to devise a unified communication-learning co-design approach for differentially private over-the-air FD. The goal is to maximize the convergence rate while meeting the transmit power and DP requirement of each WD. The contributions of this paper are threefold: 1) We consider a differentially private over-the-air FD system, where each WD uploads the local-averaged per-class model output together with the DP-preserving artificial noise to the PS per round, achieving low-dimensional disclosed signal with shared DP-preserving responsibility among WDs. 2) We theoretically analyze the over-the-air FD convergence rate with a diminishing step size under mild assumptions. In addition, we derive the closed-form expression of the required DP noise variance for WDs each round in over-the-air FD. 3) Based on the learning and DP analysis, we first obtain the short-term optimal transceiver design variables in the closed forms. Accordingly, the analytical expression of the optimal number of training rounds is derived for over-the-air FD with DP.
\section{System Model}
\subsection{FD System}
Consider an FD system \cite{OTA-FD} consisting of a single-antenna PS and \(M\) single-antenna WDs, indexed by \(i\in\mathcal{M}=\{1,2,\cdots,M\}\). Each WD \(i\) holds a local dataset \(\mathcal{B}_i\) with \(B_i\) training samples belonging to \(K\) classes, e.g., \(K=10\) for hand-writing digit classification in the MNIST database. Suppose that there are \(B_i^k\) training samples associated with the \(k\)-th class in WD \(i\), where \(\sum_{k=1}^KB_i^k=B_i\). The goal of the FD system is to learn a global model \(\bm{\theta}\) by exploiting the local datasets \(\left\{\mathcal{B}_i\right\}_{i=1}^M\) distributed among WDs. Unlike conventional FL systems where each WD's local model parameter \(\bm{\theta}_i\) or the corresponding gradient is aggregated periodically, FD systems leverage online co-distillation at each WD \(i\) under the coordination of the PS to minimize individual empirical local loss functions, i.e.,
\begin{equation}\footnotesize
    F_i\left(\bm{\theta};\left\{\mathbf{q}^k\right\}\right)=\frac{1}{B_i}\sum_{b=1}^{B_i}f\left(\bm{\theta};\mathbf{u}_i^b,v_i^b\right)+\gamma\left\|G_{\bm{\theta}}\left(\mathbf{u}_i^b\right)-\mathbf{q}^{v_i^b}\right\|^2_2,
\end{equation}
where \(\mathbf{u}_i^b\) and \(v_i^b\) denote the input feature vector and corresponding label of the \(b\)-th training sample in local dataset \(\mathcal{B}_i\), respectively. \(D\) represents the dimension of the model parameter. The output of the model, denoted by \(G_{\bm{\theta}}(\mathbf{u}_i^b)\in \mathbb{R}^K\), serves as the soft prediction based on the input \(\mathbf{u}_i^b\). To be discussed later, \(\mathbf{q}^{v_i^b}\) signifies the distilled knowledge, calculated by globally averaging the soft predictions for training samples belonging to class \(v_i^b\in\{1,\cdots,K\}\) across all WDs. As shown in (1), FD's entire empirical local loss comprises a general training loss function \(f(\cdot)\) and the \(l_2\)-norm distillation loss measuring the disparity between the model output \(G_{\bm{\theta}}\left(\cdot\right)\) and the distilled knowledge. In addition, \(\gamma\) stands for the contribution weight of the distillation loss.\par
\begin{figure}
    \centering
    \includegraphics[width=0.6\linewidth]{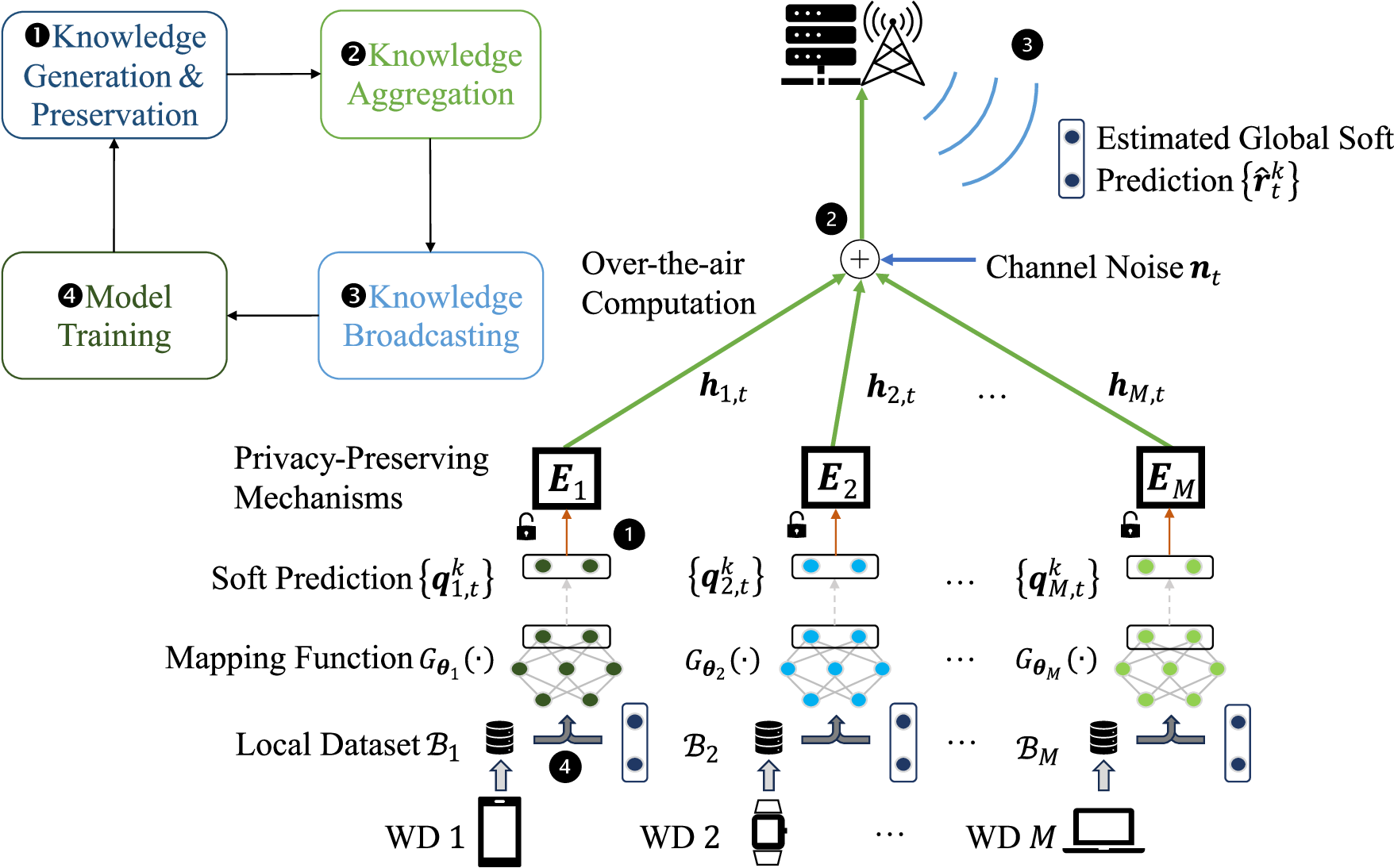}
    \caption{The considered over-the-air FD system with DP.}
\end{figure}
To cope with Problem (1), WD \(i\) leverages a gradient descent algorithm to iteratively update the model parameter \(\bm{\theta}_{i,t}\) over \(T\) training rounds. As illustrated in Fig. 1, per training round \(t\), \(1\leq t\leq T\), WD \(i\) first obtains the local knowledge \(\mathbf{q}_{i,t}^k\) of each class \(k\) by calculating the local-averaged soft prediction vector for all the associated training samples based on the current local model parameter \(\bm{\theta}_{i,t}\), i.e,
\begin{equation}
    \textstyle\mathbf{q}_{i,t}^k=\frac{1}{B_i^k}\sum_{b\in\mathcal{B}_i^k}G_{\bm{\theta}_{i,t}}\left(\mathbf{u}_i^b\right).
\end{equation}
Then, WD \(i\) sends its local knowledge \(\left\{\mathbf{q}_{i,t}^k\right\}_{k=1}^K\) to the PS via wireless channels for global knowledge aggregation. Per class \(k\), the PS tends to compute the corresponding global-averaged soft prediction vector, i.e., the distilled knowledge
    \begin{equation}
        \textstyle\mathbf{q}^k_t=\sum_{i=1}^M\frac{B_i^k}{B^k}\mathbf{q}_{i,t}^k,
    \end{equation}
    where \(B^k=\sum_{i=1}^MB_i^k\) is the total number of training samples belonging to class \(k\) across all WDs. However, due to the WDs' privacy concern, channel fading and communication noises, the PS can only obtain an estimate of \(\mathbf{q}^k_t\), denoted by \(\mathbf{\hat{r}}^k_t\). With the estimated global distilled knowledge \(\left\{\mathbf{\hat{r}}_t^k\right\}_{k=1}^K\) broadcasted to all WDs through an error-free link, WD \(i\) updates its model parameters by gradient descent, i.e.,
    \begin{equation}
    \bm{\theta}_{i,t+1}=\bm{\theta}_{i,t}-\eta_t\nabla F_i\left(\bm{\theta}_{i,t};\left\{\mathbf{\hat{r}}_t^k\right\}_{k=1}^K\right),
\end{equation}
where \(\eta_t\) is the learning rate at training round \(t\).
\subsection{DP-based Transmit Signal Design}
Although both local datasets and models are kept locally in FD, there is a potential risk of private information exposure to the PS through analyzing the transmitted knowledge \(\left\{\mathbf{q}_{i,t}^k,\forall i,k\right\}\). In this paper, we consider a honest-but-curious PS \cite{DP-FL, D-FL, User}. The DP technique is leveraged to quantify and bound the privacy leakage for each WD, given as follows \cite{DP}.\par
\textit{\textbf{Definition 1:}} Given \(\varepsilon>0,0\leq\delta\leq1\), and an arbitrary randomized mechanism \(\mathbf{E}:\mathcal{W}^n\rightarrow\mathcal{Y}\), for any two possible neighboring datasets \(\mathcal{B},\mathcal{B}'\in\mathcal{W}^n\), and any subset \(O\subseteq\mathcal{Y}\), the mechanism \(\mathbf{E}\) is \((\varepsilon,\delta)\)-differentially private if
\begin{equation}
Pr\left\{\mathbf{E}(\mathcal{B})\in O\right\}\leq e^\varepsilon Pr\left\{\mathbf{E}(\mathcal{B}')\in O\right\}+\delta,
\end{equation}
where any two neighboring datasets \(\mathcal{B}\) and \(\mathcal{B}'\) differ by a single data sample.\par
Under \((\varepsilon_i, \delta_i)\)-DP of the mechanism \(\mathbf{E}_i\) at WD \(i\), the privacy loss \(\mathcal{L}_{\mathbf{E}_i(\mathcal{B}_i)||\mathbf{E}_i(\mathcal{B}'_i)}=\ln{\frac{Pr\left\{\mathbf{E}_i(\mathcal{B}_i)\right\}}{Pr\left\{\mathbf{E}_i(\mathcal{B}'_i)\right\}}}\), defined as the log-likelihood ratio of the neighboring datasets \(\mathcal{B}_i\) and \(\mathcal{B}'_i\), satisfies
\begin{equation}
Pr\left\{\left|\mathcal{L}_{\mathbf{E}_i(\mathcal{B}_i)||\mathbf{E}_i(\mathcal{B}'_i)}\right|\leq \varepsilon_i\right\}\geq 1-\delta_i,
\end{equation}
where a smaller \(\varepsilon_i\) makes the neighboring datasets less distinguishable, yielding a higher DP level.\par
Notice that per training round \(t\), each WD \(i\) first generates the local knowledge by (2). To achieve \((\varepsilon_i,\delta_i)\)-DP, WD \(i\) adopts the Gaussian mechanism, where the Gaussian-distributed artificial noise is superposed on its local knowledge. Then, we denote the transmit signal of WD \(i\) at training round \(t\) as \(\mathbf{x}_{i,t}=\left[\mathbf{x}_{i,t}^{1^\top},\cdots,\mathbf{x}_{i,t}^{K^\top}\right]^\top\in\mathbb{C}^{K^2}\). The \(k\)-th element \(\mathbf{x}_{i,t}^k\) is given by
\begin{equation}
\mathbf{x}_{i,t}^k=P^k_{1,i,t}\sqrt{K}\mathbf{q}^k_{i,t}+P^k_{2,i,t}\mathbf{m}^k_{i,t},
\end{equation}
where each entry of the artificial noise \(\mathbf{m}^k_{i,t} \in \mathbb{R}^K\) is the independent and identically distributed (i.i.d.) Gaussian noise drawn from \(\mathcal{N}(0,1)\). \(P^k_{1,i,t}\) and \(P^k_{2,i,t}\in\mathbb{C}\) are the transmit equalization factor for the local knowledge and the DP noise of class \(k\) for WD \(i\) at training round \(t\), respectively.\par
Denote the \(d\)-th entry of \(\mathbf{x}^k_{i,t}\) as \(x^k_{i,t}\left[d\right]\). In addition, \(q^k_{i,t}[d]\) and \(m^k_{i,t}[d]\) are the \(d\)-th entry of \(\mathbf{q}^k_{i,t}\) and \(\mathbf{m}^k_{i,t}\), respectively. Then, we have\\
\begin{equation}
\begin{aligned}
    \mathbb{E}\left[\left|x^k_{i,t}[d]\right|^2\right]&=\mathbb{E}\left[\left|P^k_{1,i,t}\sqrt{K}q^k_{i,t}[d]+P^k_{2,i,t}m^k_{i,t}[d]\right|^2\right]\\
    &\leq\left|P^k_{1,i,t}\right|^2+\left|P^k_{2,i,t}\right|^2, \quad\forall d,i,k,t.
\end{aligned}
\end{equation}
Here, the inequality first follows from the independence between \(q^k_{i,t}[d]\) and the standard Gaussian \(m^k_{i,t}[d]\). Then, we approximate the second order expectation \(\mathbb{E}\left[\left|q^k_{i,t}[d]\right|^2\right]\) by the mean of the squared \(l_2\) norm of \(\mathbf{q}^k_{i,t}\), i.e., \(\frac{1}{K}\left\|\mathbf{q}^k_{i,t}\right\|_2^2\). According to the property that \(\left\|\mathbf{q}^k_{i,t}\right\|_2\) is upper bounded by the corresponding \(l_1\) norm \(\left\|\mathbf{q}^k_{i,t}\right\|_1=1\), \(\mathbb{E}\left[\left|q^k_{i,t}[d]\right|^2\right]\) is upper bounded by \(\frac{1}{K}\). In this paper, we consider that the transmit signal of each WD \(i\) is restricted by the peak transmit power \(P_i\), i.e.,
\begin{equation}
\left|P_{1,i,t}^k\right|^2+\left|P^k_{2,i,t}\right|^2\leq P_i, \quad\forall i,k,t.
\end{equation}
\subsection{Over-the-Air Knowledge Aggregation}
The superposition property of wireless
multiple-access channels enables all the WDs synchronously transmit \(\left\{\mathbf{x}_{i,t}\right\}_{i=1}^M\) entry by entry under the same radio-frequency resources. Consequently, over-the-air FD occupies a total of \(K^2\) time slots per round, yielding a much lower communication cost compared with the FL counterparts. In this paper, we assume a block fading channel model with perfect channel state information (CSI) at the PS per training round \cite{CSI}. The channel coefficients remain constant within a single FD training round and may change across different rounds. The channel coefficient between WD \(i\) and the PS at training round \(t\) is denoted as \(h_{i,t}\in\mathbb{C}\). With over-the-air computation, the received signal at the PS for time slot \(t\) is
\begin{equation}
    y_t[d]=\sum_{i=1}^Mh_{i,t}x_{i,t}[d]+n_t[d],
\end{equation}
where \(n_t[d]\sim\mathcal{N}\left(\mathbf{0},\sigma_n^2\right)\) is the additive white Gaussian noise (AWGN). The received signal corresponding to the aggregated soft prediction of class \(k\) is given by
\begin{equation}
    \mathbf{r}^k_t=\left[y_t\left[\left(k-1\right)K+1\right],\cdots,y_t\left[kK\right]\right]^\top,\forall k\in\left[1,K\right].
\end{equation}
Then, the PS adopts a linear estimator to obtain the estimated global-averaged soft prediction vector \(\mathbf{q}^k_t\) for each class \(k\), i.e.,
\begin{equation}
    \begin{aligned}
            \mathbf{\hat{r}}^k_t=\frac{\mathbf{r}^k_t}{\lambda^k_t}=\sum_{i=1}^M\frac{h_{i,t}P_{1,i,t}^k\sqrt{K}}{\lambda^k_t}\mathbf{q}_{i,t}^k+\frac{\hat{\mathbf{m}}^k_t}{\lambda^k_t},\quad \forall k,
    \end{aligned}
\end{equation}
where \(\hat{\mathbf{m}}^k_t=\sum_{i=1}^Mh_{i,t}P^k_{2,i,t}\mathbf{m}^k_{i,t}+\mathbf{n}_t^k\). Here \(\mathbf{n}_t^k=\left[n_t\left[\left(k-1\right)K+1\right],\cdots,n_t\left[kK\right]\right]^\top\) collects the i.i.d. channel noise of class \(k\). \(\lambda^k_t\) is the post-processing normalization scalar of class \(k\) in training round \(t\). Then, the estimated signal \(\left\{\mathbf{\hat{r}}^k_t\right\}_{k=1}^K\) is broadcasted to all WDs for model update in (4).
\section{Theoretical Analysis and Problem Formulation}
In this section, we first analyze the learning performance of the considered differentially private over-the-air FD system. Then, we quantitatively bound the overall privacy loss of each WD throughout \(T\) training rounds using the moment accountant technique \cite{MA}. Finally, we formulate the communication-learning co-design problem.\par
\vspace{-0.3cm}
\subsection{Convergence Analysis}
To facilitate the convergence analysis of over-the-air FD with DP, we have the following assumptions.
\begin{itemize}
    \item \textit{\textbf{Assumption 1:}} The gradient of the local loss function \(F_i(\cdot)\) is \(L_1\)-Lipschitz continuous.
    \item \textit{\textbf{Assumption 2:}} The learned model function mapping \(G_{\bm{\theta}}(\cdot)\) at each WD is \(L_2\)-Lipschitz continuous.
    \item \textit{\textbf{Assumption 3:}} The gradient norm of the loss function is uniformly bounded by a scalar \(S\).
\end{itemize}
Under Assumptions 1-3, we derive an upper bound of the expected gradient norm for each WD in the following theorem.\par
\textit{\textbf{Theorem 1:}} Suppose that Assumptions 1-3 hold and the empirical local loss function for WD \(i\) is upper bounded by \(f_{i,max}\), i.e., \(F_i\left(\bm{\theta}\right)\leq f_{i,max},\forall \bm{\theta}\). For any \(\gamma>0\) and intial learning rate \(\eta_0>0\), if \(\eta_t=\frac{\eta_0}{\sqrt{t}}\leq\frac{1}{L_1}\), then
\begin{equation}
    \mathbb{E}\left[\left\|\nabla F_i\left(\hat{\bm{\theta}}_{i,T}\right)\right\|^2_2\right]\leq8\gamma L_2S+\Omega_i\left(T,
    \left\{\mathcal{P}_t\right\}_{t=0}^{T-1}\right),
\end{equation}
where
\begin{equation}
\begin{aligned}
\Omega_i\left(T,
    \left\{\mathcal{P}_t\right\}\right)&=\frac{3f_{i,max}}{\eta_0\sqrt{T}}+\sum_{t=0}^{T-1}6\eta_0\gamma^2L_2^2L_1\left(\frac{\Phi_{1,i,t}^2+\Phi_{2,i,t}}{T^{\frac{3}{2}}}\right)\\
&\quad+\sum_{t=0}^{T-1}6\gamma\eta_0L_2\frac{\left(L_1\eta_t+1\right)\left\|\nabla F_i(\bm{\theta}_{i,t})\right\|_2\Phi_{1,i,t}}{\eta_tT^{\frac{3}{2}}}\nonumber
\end{aligned}
\end{equation}
and
\begin{equation}
    \begin{aligned}
        \Phi_{1,i,t} &= \sum_{k=1}^K\frac{B_i^k}{B_i}\left\|\sum_{i=1}^M\left(\frac{h_{i,t}P_{1,i,t}^k\sqrt{K}}{\lambda_t^k}-\frac{B_i^k}{B^k}\right)\mathbf{q}_{i,t}^k\right\|_2,\\
        \Phi_{2,i,t}&=\mathbb{E}\left[\sum_{k=1}^K\frac{B_i^k}{B_i}\left\|\frac{\hat{\mathbf{m}}^k_t}{\lambda^k_t}\right\|^2_2\right].
    \end{aligned}
\end{equation}
Here, we denote \(\Phi_{1,i,t}\) and \(\Phi_{2,i,t}\) as the abbreviations of \(\Phi_{1,i,t}\left(\left\{P_{1,i,t}^k,\lambda_t^k,\forall i,k\right\}\right)\) and \(\Phi_{2,i,t}\left(\left\{P_{2,i,t}^k,\lambda_t^k,\forall i,k\right\}\right)\), respectively. \(\mathcal{P}_t=\left\{P_{1,i,t}^k,\lambda_t^k,P^k_{2,i,t},\forall i,k\right\}\) is the transceiver design at round \(t\). Moreover, \(\hat{\bm{\theta}}_{i,T}\) is randomly chosen from \(\left\{\bm{\theta}_{i,t},\forall t\right\}\) with probability \(Pr\left\{\hat{\bm{\theta}}_{i,T}=\bm{\theta}_{i,t}\right\}=\frac{1/\eta_t}{\sum_{t=0}^{T-1}1/\eta_t}\).\par
\begin{proof}
    See Appendix A.
\end{proof}
We observe that the error function \(\Phi_{1,i,t}\left(\cdot\right)\) is due to the possible signal misalignment caused by the channel fading. \(\Phi_{2,i,t}\left(\cdot\right)\) arises from the perturbation of the DP-preserving and Gaussian channel noise.
\subsection{DP Analysis}
The artificial noise at each WD for the DP preservation is related to the sensitivity of the noise-free revealed signal of the WD, i.e., \(\mathbf{q}_{i,t}=\left[\hat{\mathbf{q}}_{i,t}^{1^\top},\cdots,\hat{\mathbf{q}}_{i,t}^{K^\top}\right]^\top\in\mathbb{C}^{K^2}\). Specifically, \(\hat{\mathbf{q}}_{i,t}^k=h_{i,t}P^k_{1,i,t}\sqrt{K}\mathbf{q}^k_{i,t}\). Associated with two adjacent datasets \(\mathcal{B}_i\) and \(\mathcal{B}'_i\) for WD \(i\), the disclosed signals of class \(k\) at training round \(t\) are \(\hat{\mathbf{q}}^k_{i,t}\left(\mathcal{B}_i\right)\) and \(\hat{\mathbf{q}}^k_{i,t}\left(\mathcal{B}'_i\right)\), respectively. The corresponding \(l_2\) sensitivity is then calculated as\\
\begin{equation}
\Delta^k_{i,t}=\max_{\mathcal{B}_i,\mathcal{B}'_i}\left\|\hat{\mathbf{q}}^k_{i,t}\left(\mathcal{B}_i\right)-\hat{\mathbf{q}}^k_{i,t}\left(\mathcal{B}'_i\right)\right\|_2.
\end{equation}
Here, the \(l_2\) sensitivity measures the maximum impact on the leaked information caused by altering a single sample in the input dataset.\par
Since the elements of \(\mathbf{q}_{i,t}^k\) sum to \(1\) for any \(k\) due to the softmax layer, the collection of all possible \(\mathbf{q}_{i,t}^k\) forms a probability simplex \(\Sigma_{i,t}^k=\left\{\mathbf{q}_{i,t}^k\in\mathbb{R}^K\middle|\mathbf{1}^\top\mathbf{q}_{i,t}^k=1,\mathbf{q}_{i,t}^k\geq\mathbf{0}\right\}\), where \(\mathbf{1}\) is $K$-dimensional all-one vector. Therefore, bounding the sensitivity in (15) is equivalent to finding the maximum Euclidean distance between any two points in a probability simplex.\par
\textit{\textbf{Lemma 1:}} The \(l_2\) sensitivity in (15) is upper bounded by
\begin{equation}
\Delta^k_{i,t}\leq\sqrt{2K}\left|h_{i,t}P_{1,i,t}^k\right|/B_i,\quad\forall i,k,t.
\end{equation}
\begin{proof}
See Appendix B.
\end{proof}
By bounding the sensitivity of each class per round in Lemma 1 and accumulating \(T\)-round privacy losses in the considered FD system \cite{MA}, we determine the variance of the artificial noise (i.e., find the
transmit power \(P^k_{2,i,t}, \forall i, k, t\)) to fulfill (\(\varepsilon_i, \delta_i\))-DP of each WD in the following theorem.\par

\textit{\textbf{Theorem 2:}} To achieve \((\varepsilon_i,\delta_i)\)-DP of WD \(i\) after \(T\) training rounds, the following inequality must hold:
\begin{equation}
\sum_{j=1}^M\left|h_{j,t}P^k_{2,j,t}\right|^2+\sigma_n^2\geq\max_{i\in\mathcal{M}}4TK\left|h_{i,t}P_{1,i,t}^k\right|^2\rho_i,\forall k,t,
\end{equation}
where \(\rho_i = \frac{\ln{\frac{1}{\delta_i}}}{B_i^2\varepsilon_i^2}\) measures the stringency of WD \(i\)'s privacy requirement. We observe that a larger training round \(T\), class number \(K\) and privacy requirement \(\rho_i\) result in a higher variance of privacy-preserving artificial noises.
\subsection{Problem Formulation}
With Theorem 1 and 2, we formulate the communication-learning co-design problem by minimizing the upper bound of the gradient norm in (13) of all the WDs. Our goal is to optimize the number of training rounds \(T\) and the transceiver design \(\mathcal{P}_t\) at each round subject to the peak transmit power constraint in (9) and the privacy constraint in (17), i.e.,
\begin{equation}
        \textstyle\mbox{(P1)}\quad \underset{T\in\mathbb{N},\left\{\mathcal{P}_t\right\}}{\text{min}}\;\sum_{i=1}^M\Omega_i\left(T,
    \left\{\mathcal{P}_t\right\}_{t=0}^{T-1}\right)\quad\text{s.t.} \;(9),(17).\nonumber
\end{equation}
 The challenge in solving Problem (P1) lies in the strong coupling of the continuous transmit power allocation decisions for local distilled knowledge and DP noise transmission per round, further mixed with the integer training rounds \(T\) at a larger timescale.
\section{Communication-Learning Co-design}
To tackle Problem (P1), we first obtain the closed-form solutions of the transceiver design per round given the number of training rounds \(T\). Based on the per-round decisions, the large time-scale optimization over \(T\) is derived.\par
\textit{\textbf{Proposition 1:}} Per training round \(t\), given the number of training rounds \(T\), the optimal transmit power for soft prediction uploading is\\
\begin{equation}
    P_{1,i,t}^{k^\ast} = \frac{B^k_i\lambda_t^k\overline{h}_{i,t}}{B^k\sqrt{K}\left|h_{i,t}\right|^2},\quad\forall i,k.
\end{equation}
Moreover, the optimal transmit power for the DP noise \(\left\{P^k_{2,i,t},\forall i\right\}\) and post-processing scalar \(\left\{\lambda_t^k,\forall k\right\}\) at round \(t\) are obtained in the following two cases:
\begin{itemize}
    \item If \(T\leq\frac{\sigma_n^2}{4K\left(\min_{i\in\mathcal{M}}|h_{i,t}|^2P_i\right)\max_{i\in\mathcal{M}}\rho_i}\),
    \begin{equation}
        \begin{aligned}
            \lambda_t^{k^\ast}&=\min_{i\in\mathcal{M}}\frac{B^k\sqrt{K}\left|h_{i,t}\right|\sqrt{P_i}}{B_i^k},\quad\forall k\\
            P_{2,i,t}^{k^\ast}&=0,\quad\forall i,k.
        \end{aligned}
    \end{equation}
    \item Otherwise, the optimal \(\left\{P^k_{2,i,t},\lambda_t^k\right\}\) are any solution to
    \begin{equation}
    \sum_{j=1}^M\left|h_{j,t}P^{k^\ast}_{2,j,t}\right|^2+\sigma_n^2=4T\left(\lambda_t^{k^\ast}\right)^2\max_{i\in\mathcal{M}}\left(\frac{B_i^k}{B^k}\right)^2\rho_i,\;\forall k.
    \end{equation}
\end{itemize}
\begin{proof}
    See Appendix C.
\end{proof}
From the above proposition, we can observe that the transmit power for soft prediction \(P_{1,i,t}^k\) increases when WD \(i\) holds a larger portion of training samples of class \(k\). In addition, one can harness the inherent channel noise to preserve DP with a sufficiently small training rounds \(T\) or privacy requirements.\par
\textit{\textbf{Proposition 2:}} Given the optimal transceiver design per round, the optimal number of training rounds is given by
\begin{equation}
    T^\ast=\lfloor\frac{\frac{3}{M}\sum_{i=1}^Mf_{i,max}}{24\eta_0^2\gamma^2L_2^2L_1\max_{i\in\mathcal{M}}\rho_i\sum_{k=1}^K\left(\frac{B_i^k}{B^k}\right)^2}\rceil.
\end{equation}
\begin{proof}
    See Appendix D.
\end{proof}
\(\lfloor\cdot\rceil\) rounds the input to the nearest integer. WDs tend to train the learning models for more FD rounds if each of them has a higher loss value \(f_{i,max}\). On the other hand, the considered FD system prefers shorter training rounds \(T\) with a more stringent DP requirement \(\max_{i\in\mathcal{M}}\rho_i\) among WDs. Regrading the impact of data distributions, more training rounds will be conducted with balanced per-class datasets among WDs. If one WD holds most of the data samples from the same label, the number of training round \(T\) decreases since the local distilled knowledge from that specific WD would leak more private information, requiring a larger DP noise variance each round.
\section{Numerical Results}
In this section, we perform extensive simulations to evaluate and compare the performance of the proposed differentially-private over-the-air FD.\par
We assume a total of \(M=50\) WDs in the considered FD system. The channel coefficient \(h_{i,t}\) is modelled by multiplying the small-scale fading coefficient on the large-scale path loss. The path loss is given by \(\left(\frac{3\times10^8}{4\pi f_c d_i}\right)^{PL}\), where \(f_c=915\) MHz is the carrier frequency. \(d_i\) is the distance between WD \(i\) and the PS, which is distributed uniformly from \(100\) to \(500\) m. \(PL\) is the path loss exponent. The small-scale fading coefficients are i.i.d. standard Gaussian random variables across training rounds. All the methods are evaluated on the image classification task over the MNIST database, which consists of 60000 training samples and 10000 testing images from \(K=10\) classes. WDs collaboratively train a convolutional neural network (CNN), which consists of two convolution layers with the max pooling and a ReLU activation function, two fully connected layers and a softmax layer. The number of model parameters altogether is \(D=21680\). The learning rate decaying factor is \(1/\sqrt{t}\). For the FD approaches, the DP requirement \(\varepsilon_i\) is drawn from the uniform distribution from \(0.001\) to \(0.1\), while \(\delta_i\) is from \(10^{-11}\) to \(10^{-9}\). Each WD \(i\) has a peak transmit power \(P_i=1\) mW. The variance of the channel noise is \(\sigma_n^2=10^{-8}\) W and \(PL=3\). The initial learning rate \(\eta_0\) is 0.01.\par
For performance comparison, we consider the FedSGD algorithm on the error-free channel as the baseline. In addition, the proposed method is benchmarked by the approach in \cite{D-FL}, where the number of training rounds \(T\) is optimized for the over-the-air FL system with DP. Moreover, we also simulate the learning tasks with constant \(T=400\) in the FD and the FL systems, respectively, named FD/FL Constant \(T\).\par
\begin{figure}
\centering
    \subfigure[Effective noise variance \(\Phi_{2,i,t}\)]{\begin{minipage}[b]{0.49\linewidth}
    \centering
            \includegraphics[width=\linewidth]{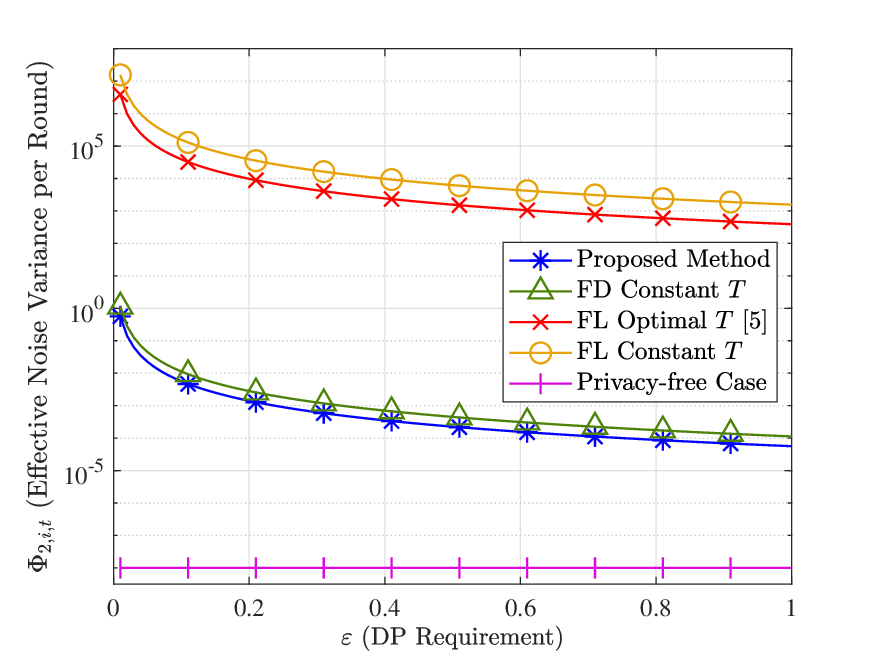}
        \end{minipage}}
        \subfigure[Average testing accuracies]{\begin{minipage}[b]{0.49\linewidth}
        \centering
            \includegraphics[width=\linewidth]{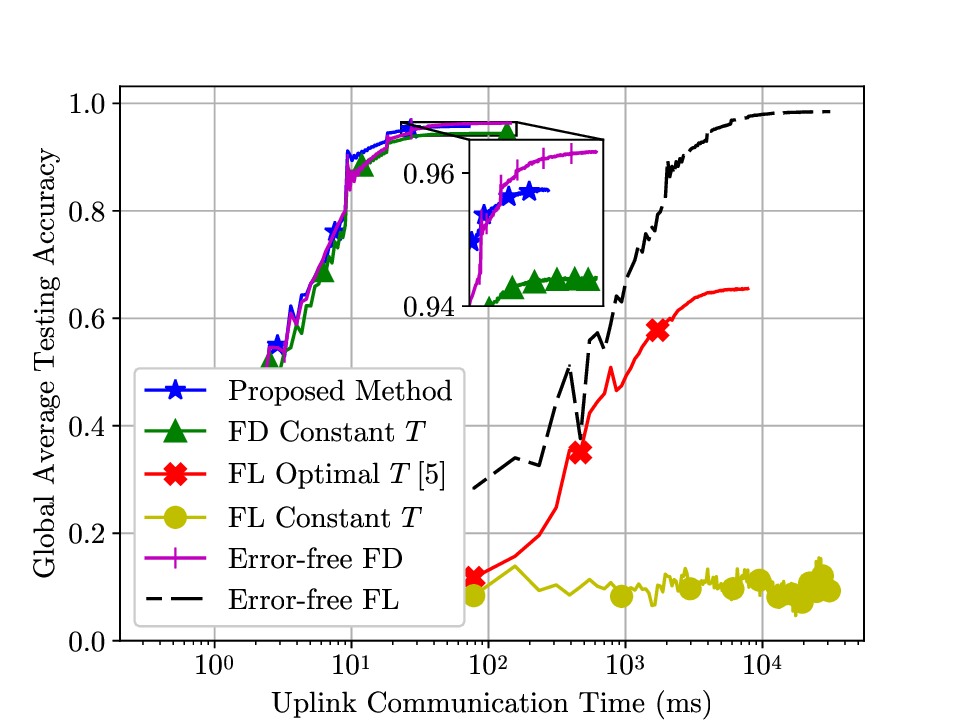}
        \end{minipage}}
    \caption{(a) The effective variance \(\Phi_{2,i,t}\) of aggregated DP-preserving and channel noises at the PS per training round versus the DP requirement \(\varepsilon\), and (b) the average testing accuracies versus the uplink communication time.}
\end{figure}
In Fig. 2 (a), we track and plot \(\Phi_{2,i,t}\), the amount of aggregated noises at the PS per training round, versus the DP requirement \(\varepsilon\) of WDs. We choose \(\varepsilon_i=\varepsilon\), \(\delta_i=10^{-11},\forall i\) for both FL and FD systems. Since the sensitivity of
the privacy-preserved signal is proportional to its dimension, model aggregation in FL methods suffers from a much larger DP noise than that in FD. Accordingly, better wireless channel conditions are considered for the FL baselines. Specifically, Each WD \(i\) has a peak transmit power \(P_i=10\) mW. The variance of the channel noise is \(\sigma_n^2=10^{-10}\) W and \(PL=2\). The initial learning rate \(\eta_0\) is 0.001. We observe that in spite of better channel conditions, the FL methods with both fixed and optimized \(T\) suffer from significantly larger DP-preserving noises than those in the FD system. This phenomenon signifies the superiority of the considered over-the-air FD system when each WD requires the DP preservation during the distributed learning process, thanks to the largely reduced dimension of the disclosed information per round. Moreover, the amount of aggregated noises is further reduced with the training rounds \(T\) optimized via the proposed algorithm for differentially private over-the-air FD.\par
Fig. 2 (b) depicts the testing accuracies with respect to the uplink communication time. We consider that each time slot of the transmitted signal in the uplink occupies 3.6 \(\mu\)s, following the orthogonal frequency-division multiplexing (OFDM) standard \cite{OFDM}. Due to the high-dimensional disclosed signal in FL approaches, we consider milder DP requirements for the ease of performance comparison. Specifically, \(\varepsilon_i\) is drawn from the uniform distribution from \(1\) to \(100\), while \(\delta_i\) is from \(10^{-4}\) to \(10^{-3}\) for the FL benchmarks. Firstly, it is observed that with the optimal training rounds \(T\), the proposed method achieves a higher testing accuracy compared to the constant \(T\) FD benchmark, demonstrating the necessity of the training rounds optimization in over-the-air FD with DP. Then, we find that the proposed method approaches the error-free FD and achieves the highest testing accuracy with the minimum uplink communication time compared to the FL alternatives, despite undergoing more stringent DP requirements. This indicates that the proposed differentially-private over-the-air FD system not only significantly reduces the communication overhead, but is also more robust to the privacy-preserving artificial noise compared to conventional over-the-air FL with DP.
\section{Conclusions}
This paper investigated the communication-learning co-design for differentially private over-the-air FD. With the theoretical analysis of the FD learning convergence and privacy loss for each WD, we formulated the co-design problem by maximizing the convergence rate while meeting the power and privacy requirements of WDs. The closed-form solutions were derived for the optimal transmit power, estimator for over-the-air aggregation, and the optimal number of FD rounds. Simulation results demonstrated that the proposed approach not only reduces communication overheads, but is also more robust against DP noises compared with the conventional FL.
{\appendices
\section{Proof of Theorem 1}
\begin{proof}
We first bound the expected difference of the empirical local loss function of WD \(i\) between iterations \(t\) and \(t+1\), i.e.,
\begin{equation}
        \mathbb{E}\left[F_i\left(\bm{\theta}_{i,t+1}\right)-F_i\left(\bm{\theta}_{i,t}\right)\right]=B=B_1+B_2,
\end{equation}
where
\begin{equation}
    \begin{aligned}
        B_1&=\mathbb{E}\left[F_i\left(\bm{\theta}_{i,t+1};\left\{\mathbf{q}_{t+1}^k\right\}\right)-F_i\left(\bm{\theta}_{i,t+1};\left\{\mathbf{q}_t^k\right\}\right)\right],\\
        B_2&=\mathbb{E}\left[F_i\left(\bm{\theta}_{i,t+1};\left\{\mathbf{q}_t^k\right\}\right)-F_i\left(\bm{\theta}_{i,t};\left\{\mathbf{q}_t^k\right\}\right)\right].
    \end{aligned}
\end{equation}
By Assumption 1 and Eqn. (4), we have
\begin{equation}
\begin{aligned}
    B_2&\leq\underbrace{-\eta_t\nabla F_i\left(\bm{\theta}_{i,t}\right)^\top\mathbb{E}\left[\nabla F_i\left(\bm{\theta}_{i,t};\left\{\hat{\mathbf{r}}_t^k\right\}\right)\right]}_{C_1}\\
    &\quad+\frac{L_1\eta_t^2}{2}\underbrace{\mathbb{E}\left[\left\|\nabla F_i\left(\bm{\theta}_{i,t};\left\{\hat{\mathbf{r}}_t^k\right\}\right)\right\|^2_2\right]}_{C_2}.
\end{aligned}
\end{equation}
According to (1)-(3),
\begin{equation}
\begin{aligned}
    B_1&=\frac{\gamma}{B_i}\sum_{b=1}^{B_i}\left(\left\|G_{\bm{\theta}_{i,t+1}}\left(\mathbf{u}_i^b\right)-\mathbf{q}^{v_i^b}_{t+1}\right\|^2_2-\left\|G_{\bm{\theta}_{i,t+1}}\left(\mathbf{u}_i^b\right)-\right.\right.\\
    &\qquad\qquad\quad\left.\left.\mathbf{q}^{v_i^b}_t\right\|^2_2\right)\nonumber
    \end{aligned}
    \end{equation}
    \begin{equation}
    \begin{aligned}
    &= \frac{\gamma}{B_i}\sum_{b=1}^{B_i}\left(2G_{\bm{\theta}_{i,t+1}}\left(\mathbf{u}_i^b\right)^\top\mathbf{q}^{v_i^b}_t-2G_{\bm{\theta}_{i,t+1}}\left(\mathbf{u}_i^b\right)^\top\mathbf{q}^{v_i^b}_{t+1}\right.\\
    &\qquad\qquad\quad\left.+\left\|\mathbf{q}^{v_i^b}_{t+1}\right\|^2_2-\left\|\mathbf{q}^{v_i^b}_t\right\|^2_2\right)\\
    &=\frac{\gamma}{B_i}\sum_{b=1}^{B_i}\left(\mathbf{q}^{v_i^b}_{t+1}+\mathbf{q}^{v_i^b}_t-2G_{\bm{\theta}_{i,t+1}}\left(\mathbf{u}_i^b\right)\right)^\top\left(\mathbf{q}^{v_i^b}_{t+1}-\mathbf{q}^{v_i^b}_t\right)\\
    &\stackrel{(a)}{\leq}\frac{\gamma}{B_i}\sum_{b=1}^{B_i}\left\|\mathbf{q}^{v_i^b}_{t+1}+\mathbf{q}^{v_i^b}_t-2G_{\bm{\theta}_{i,t+1}}\left(\mathbf{u}_i^b\right)\right\|_2\left\|\mathbf{q}^{v_i^b}_{t+1}-\mathbf{q}^{v_i^b}_t\right\|_2\\
    &\stackrel{(b)}{\leq}\frac{\gamma}{B_i}\sum_{b=1}^{B_i}\left(\left\|\mathbf{q}^{v_i^b}_{t+1}\right\|_2+\left\|\mathbf{q}^{v_i^b}_t\right\|_2+\left\|2G_{\bm{\theta}_{i,t+1}}\left(\mathbf{u}_i^b\right)\right\|_2\right)\\
    &\qquad\qquad\quad\times\left\|\mathbf{q}^{v_i^b}_{t+1}-\mathbf{q}^{v_i^b}_t\right\|_2\\
    &\stackrel{(c)}{\leq}\frac{\gamma}{B_i}\sum_{k=1}^{K}B_i^k4\left\|\mathbf{q}^k_{t+1}-\mathbf{q}^k_t\right\|_2\\
    &=\frac{4\gamma}{B_i}\sum_{k=1}^{K}B_i^k\left\|\frac{1}{B^k}\sum_{i=1}^M\sum_{b\in\mathcal{B}_i^k}\left(G_{\bm{\theta}_{i,t+1}}\left(\mathbf{u}_i^b\right)-G_{\bm{\theta}_{i,t}}\left(\mathbf{u}_i^b\right)\right)\right\|_2\\
    &\stackrel{(d)}{\leq}\frac{4\gamma}{B_i}\sum_{k=1}^{K}\frac{B_i^k}{B^k}\sum_{i=1}^M\sum_{b\in\mathcal{B}_i^k}L_2\left\|\bm{\theta}_{i,t+1}-\bm{\theta}_{i,t}\right\|_2\\
        &=\frac{4\gamma}{B_i}\sum_{k=1}^{K}\frac{B_i^k}{B^k}\sum_{i=1}^M\sum_{b\in\mathcal{B}_i^k}L_2\left\|-\eta_t\nabla F_i\left(\bm{\theta}_{i,t};\left\{\hat{\mathbf{r}}_t^k\right\}\right)\right\|_2\\
    &\stackrel{(e)}{\leq}4\gamma L_2\eta_tS,
    \end{aligned}
\end{equation}
where (\(a\)) is due to Cauchy–Schwarz Inequality. Inequality (\(b\)) follows from the triangle inequality. For the inequality (\(c\)), we first see that the \(l_1\) norms of \(\mathbf{q}^{v_i^b}_{t+1}\), \(\mathbf{q}^{v_i^b}_t\) and \(G_{\bm{\theta}_{i,t+1}}\left(\mathbf{u}_i^b\right)\) are equal to 1 since they are the average of probability vectors (outputs of the softmax layer) whose entries sum to 1. Then according to the property of \(l_1\), \(l_2\) norms, for any vector \(\mathbf{x}\in\mathbb{R}^d\), we have \(\left\|\mathbf{x}\right\|_2\leq\left\|\mathbf{x}\right\|_1\), and the \(l_2\) norms of \(\mathbf{q}^{v_i^b}_{t+1}\), \(\mathbf{q}^{v_i^b}_t\) and \(G_{\bm{\theta}_{i,t+1}}\left(\mathbf{u}_i^b\right)\) are bounded by 1. Subsequently, we interchange the summation of sample indices \(b\) with class indices \(k\). Inequality (\(d\)) is due to Jensen's Inequality and Assumption 2. Inequality (\(e\)) follows from Assumption 3.\par
Next, we bound \(C_1\) and \(C_2\). By (1) and chain rule, we have
\begin{equation}
\begin{aligned}
    \nabla F_i\left(\bm{\theta}_{i,t};\left\{\hat{\mathbf{r}}_t^k\right\}\right)&=\nabla F_i\left(\bm{\theta}_{i,t}\right)+\frac{2\gamma}{B_i}\sum_{b=1}^{B_i}\frac{\partial G_{\bm{\theta}_{i,t}}\left(\mathbf{u}_i^b\right)}{\partial\bm{\theta}_{i,t}}\\
    &\quad\times\left(G_{\bm{\theta}_{i,t}}\left(\mathbf{u}_i^b\right)-\hat{\mathbf{r}}_t^{v_i^b}\right)\\
    &=\nabla F_i\left(\bm{\theta}_{i,t}\right)+\frac{2\gamma}{B_i}\sum_{b=1}^{B_i}\frac{\partial G_{\bm{\theta}_{i,t}}\left(\mathbf{u}_i^b\right)}{\partial\bm{\theta}_{i,t}}\\
    &\quad\times\left(G_{\bm{\theta}_{i,t}}\left(\mathbf{u}_i^b\right)-\hat{\mathbf{r}}_t^{v_i^b}+\mathbf{q}_t^{v_i^b}-\mathbf{q}_t^{v_i^b}\right)\\
    &=\nabla F_i\left(\bm{\theta}_{i,t}\right)+\frac{2\gamma}{B_i}\sum_{b=1}^{B_i}\frac{\partial G_{\bm{\theta}_{i,t}}\left(\mathbf{u}_i^b\right)}{\partial\bm{\theta}_{i,t}}\\
    &\quad\times\left(\mathbf{q}_t^{v_i^b}-\hat{\mathbf{r}}_t^{v_i^b}\right).
\end{aligned}
\end{equation}
By plugging (26) into \(C_1\), we get
\begin{equation}
    \begin{aligned}
        C_1&=-\eta_t\left\|\nabla F_i\left(\bm{\theta}_{i,t}\right)\right\|^2_2-\frac{2\eta_t\gamma}{B_i}\sum_{b=1}^{B_i}\nabla F_i\left(\bm{\theta}_{i,t}\right)^\top\frac{\partial G_{\bm{\theta}_{i,t}}\left(\mathbf{u}_i^b\right)}{\partial\bm{\theta}_{i,t}}\\
        &\quad\times\left(\mathbf{q}_t^{v_i^b}-\mathbb{E}\left[\hat{\mathbf{r}}_t^{v_i^b}\right]\right)\\
        &\stackrel{(f)}{\leq}-\eta_t\left\|\nabla F_i\left(\bm{\theta}_{i,t}\right)\right\|^2_2+2\gamma L_2\eta_t\left\|\nabla F_i\left(\bm{\theta}_{i,t}\right)\right\|_2\sum_{k=1}^K\frac{B_i^k}{B_i}\\
        &\quad\times\left\|\mathbf{q}_t^k-\mathbb{E}\left[\hat{\mathbf{r}}_t^k\right]\right\|_2\\
        &=-\eta_t\left\|\nabla F_i\left(\bm{\theta}_{i,t}\right)\right\|^2_2+2\gamma L_2\eta_t\left\|\nabla F_i\left(\bm{\theta}_{i,t}\right)\right\|_2\Phi_{1,i,t},
    \end{aligned}
\end{equation}
where (\(f\)) is due to the Cauchy–Schwarz Inequality that
\begin{equation}
\begin{aligned}
    \nabla F_i\left(\bm{\theta}_{i,t}\right)^\top\frac{\partial G_{\bm{\theta}_{i,t}}\left(\mathbf{u}_i^b\right)}{\partial\bm{\theta}_{i,t}}\left(\mathbf{q}_t^{v_i^b}-\mathbb{E}\left[\hat{\mathbf{r}}_t^{v_i^b}\right]\right)\geq&\\
    -\left\|\nabla F_i\left(\bm{\theta}_{i,t}\right)\right\|_2\left\|\frac{\partial G_{\bm{\theta}_{i,t}}\left(\mathbf{u}_i^b\right)}{\partial\bm{\theta}_{i,t}}\right\|_2\left\|\mathbf{q}_t^{v_i^b}-\mathbb{E}\left[\hat{\mathbf{r}}_t^{v_i^b}\right]\right\|_2&.
\end{aligned}
\end{equation}
The inequality is also due to the property of the matrix \(2-\)norm that for any matrix \(\mathbf{A}\in\mathbb{R}^{M\times N}\) and vector \(\mathbf{x}\in\mathbb{R}^N\), \(\left\|\mathbf{A}\mathbf{x}\right\|_2\leq\left\|\mathbf{A}\right\|_2\left\|\mathbf{x}\right\|_2\). Notice that by Assumption 2, we can bound the norm of the partial derivative of the learned model function \(G_{\bm{\theta}_{i,t}}\left(\cdot\right)\) by the Lipschitz constant \(L_2\). The last equality follows from the zero mean of \(\hat{\mathbf{m}}_t^k\) and the definition of \(\Phi_{1,i,t}\) in (14).\par
Similarly, with Eqn. (26), we have
\begin{equation}
    \begin{aligned}
        C_2 &= \left\|\nabla F_i\left(\bm{\theta}_{i,t}\right)\right\|^2_2+\frac{4\gamma}{B_i}\sum_{b=1}^{B_i}\nabla F_i\left(\bm{\theta}_{i,t}\right)^\top\frac{\partial G_{\bm{\theta}_{i,t}}\left(\mathbf{u}_i^b\right)}{\partial\bm{\theta}_{i,t}}\left(\mathbf{q}_t^{v_i^b}-\right.\\
        &\quad\left.\mathbb{E}\left[\hat{\mathbf{r}}_t^{v_i^b}\right]\right)+\mathbb{E}\left[\left\|\frac{2\gamma}{B_i}\sum_{b=1}^{B_i}\frac{\partial G_{\bm{\theta}_{i,t}}\left(\mathbf{u}_i^b\right)}{\partial\bm{\theta}_{i,t}}\left(\mathbf{q}_t^{v_i^b}-\hat{\mathbf{r}}_t^{v_i^b}\right)\right\|^2_2\right]\\
                &\stackrel{(g)}{\leq}\left\|\nabla F_i\left(\bm{\theta}_{i,t}\right)\right\|^2_2+4\gamma L_2\left\|\nabla F_i\left(\bm{\theta}_{i,t}\right)\right\|_2\Phi_{1,i,t}\\
        &\quad+\mathbb{E}\left[\left\|\frac{2\gamma}{B_i}\sum_{b=1}^{B_i}\frac{\partial G_{\bm{\theta}_{i,t}}\left(\mathbf{u}_i^b\right)}{\partial\bm{\theta}_{i,t}}\left(\mathbf{q}_t^{v_i^b}-\hat{\mathbf{r}}_t^{v_i^b}\right)\right\|^2_2\right]\\
        &\stackrel{(h)}{\leq}\left\|\nabla F_i\left(\bm{\theta}_{i,t}\right)\right\|^2_2+4\gamma L_2\left\|\nabla F_i\left(\bm{\theta}_{i,t}\right)\right\|_2\Phi_{1,i,t}\\
        &\quad+4\gamma^2L_2^2\left(\Phi_{1,i,t}^2+\Phi_{2,i,t}\right),
            \end{aligned}
        \end{equation}
where (\(g\)) follows from (27). The inequality (\(h\)) is due to Jensen's Inequality as well as Assumption 2. By the independence and the zero mean of \(\hat{\mathbf{m}}_t^k\), \(\mathbb{E}\left[\sum_{k=1}^{K}\frac{B_i^k}{B_i}\left\|\mathbf{q}_t^{v_i^b}-\hat{\mathbf{r}}_t^{v_i^b}\right\|^2_2\right]\) can be decomposed into \(\Phi_{1,i,t}^2+\Phi_{2,i,t}\).\par
Plug (24), (25), (27) and (29) back to (22), we get
\begin{equation}
    \begin{aligned}
        B&\leq \left(\frac{\eta_t^2L_1}{2}-\eta_t\right)\left\|\nabla F_i\left(\bm{\theta}_{i,t}\right)\right\|^2_2+L_1L_2^2\eta_t^2\left(\Phi_{1,i,t}^2+\Phi_{2,i,t}\right)\\
        &\quad\times2\gamma^2+2\gamma L_2\eta_t\left(L_1\eta_t+1\right)\left\|\nabla F_i\left(\bm{\theta}_{i,t}\right)\right\|_2\Phi_{1,i,t}\\
        &\quad+4\gamma L_2\eta_tS\\
        &\leq -\frac{\eta_t}{2}\left\|\nabla F_i\left(\bm{\theta}_{i,t}\right)\right\|^2_2+2\gamma L_2\eta_t\left(L_1\eta_t+1\right)\left\|\nabla F_i\left(\bm{\theta}_{i,t}\right)\right\|_2\\
    &\quad\times\Phi_{1,i,t}+2\gamma^2L_1L_2^2\eta_t^2\left(\Phi_{1,i,t}^2+\Phi_{2,i,t}\right)+4\gamma L_2\eta_tS.
    \end{aligned}
\end{equation}
The last inequality is due to \(-\eta_t+\frac{\eta_t^2L_1}{2}\leq -\frac{\eta_t}{2}\) with \(\eta_t\leq \frac{1}{L_1}\). We rearrange the terms, divide both sides by \(\frac{\eta_t^2}{2}\) and sum over \(t=0\) to \(T-1\) to obtain
\begin{equation}
    \begin{aligned}
        \underbrace{\sum_{t=0}^{T-1}\frac{1}{\eta_t}\left\|\nabla F_i\left(\bm{\theta}_{i,t}\right)\right\|^2_2}_{C_3}&\leq \sum_{t=0}^{T-1}\frac{2}{\eta_t^2}\mathbb{E}\left[F_i\left(\bm{\theta}_{i,t}\right)-F_i\left(\bm{\theta}_{i,t+1}\right)\right]\\
        &\quad+\sum_{t=0}^{T-1}\frac{4}{\eta_t}\gamma L_2\left(L_1\eta_t+1\right)\left\|\nabla F_i\left(\bm{\theta}_{i,t}\right)\right\|_2\\
        &\quad\times\Phi_{1,i,t}+\sum_{t=0}^{T-1}\frac{8}{\eta_t}\gamma L_2S\\
        &\quad+\sum_{t=0}^{T-1}4\gamma^2L_1L_2^2\left(\Phi_{1,i,t}^2+\Phi_{2,i,t}\right).
        \end{aligned}
\end{equation}
According to the proof of Theorem 3.5 in \cite{ref12},
\begin{equation}
    \begin{aligned}
        C_3&\leq\frac{2}{\eta_0^2}\left[F_i\left(\bm{\theta}_{i,0}\right)+\sum_{t=1}^{T-1}\left(t-\left(t-1\right)\right)\mathbb{E}\left[F_i\left(\bm{\theta}_{i,t}\right)\right]\right]\\
        &\quad+\sum_{t=0}^{T-1}\frac{8}{\eta_t}\gamma L_2S+\sum_{t=0}^{T-1}4\gamma^2L_1L_2^2\left(\Phi_{1,i,t}^2+\Phi_{2,i,t}\right)\\
        &\quad+\sum_{t=0}^{T-1}\frac{4}{\eta_t}\gamma L_2\left(L_1\eta_t+1\right)\left\|\nabla F_i\left(\bm{\theta}_{i,t}\right)\right\|_2\Phi_{1,i,t}\\
        &\leq\frac{2Tf_{i,max}}{\eta_0^2}+\sum_{t=0}^{T-1}\frac{4}{\eta_t}\gamma L_2\left(L_1\eta_t+1\right)\left\|\nabla F_i\left(\bm{\theta}_{i,t}\right)\right\|_2\Phi_{1,i,t}\\
        &\quad+\sum_{t=0}^{T-1}\frac{8}{\eta_t}\gamma L_2S+\sum_{t=0}^{T-1}4\gamma^2L_1L_2^2\left(\Phi_{1,i,t}^2+\Phi_{2,i,t}\right).
    \end{aligned}
\end{equation}
Recall that \(\hat{\bm{\theta}}_{i,T}\) is randomly chosen from \(\left\{\bm{\theta}_{i,t},\forall t\right\}\) at all the previous iterations with probability \(Pr\left\{\hat{\bm{\theta}}_{i,T}=\bm{\theta}_{i,t}\right\}=\frac{1/\eta_t}{\sum_{t=0}^{T-1}1/\eta_t}\). We divide both sides by \(\sum_{t=0}^{T-1}\frac{1}{\eta_t}\), which gives
\begin{equation}
\begin{aligned}
        \mathbb{E}\left[\left\|\nabla F_i\left(\hat{\bm{\theta}}_{i,T}\right)\right\|^2_2\right]&\leq\frac{3f_{i,max}}{\eta_0\sqrt{T}}+\sum_{t=0}^{T-1}\frac{6\gamma\eta_0L_2\left(L_1\eta_t+1\right)}{\eta_t}\\
                &\quad\times\frac{\left\|\nabla F_i(\bm{\theta}_{i,t})\right\|_2\Phi_{1,i,t}}{T^{\frac{3}{2}}}+8\gamma L_2S\\
        &\quad+\sum_{t=0}^{T-1}6\eta_0\gamma^2L_2^2L_1\left(\frac{\Phi_{1,i,t}^2+\Phi_{2,i,t}}{T^{\frac{3}{2}}}\right)
            \end{aligned}
        \end{equation}
where the inequality holds because \(\sum_{t=0}^{T-1}\frac{1}{\eta_t}\geq \frac{1}{\eta_0}\int_{t=0}^T\sqrt{t}\,dt=\frac{2}{3\eta_0}T^{\frac{3}{2}}\). We summarize all the terms related to the training rounds \(T\) and per-round transceiver design \(\mathcal{P}_t\) in (33) into the convergence gap function \(\Omega_i\left(T,\left\{\mathcal{P}_t\right\}_{t=0}^{T-1}\right)\). The proof of Theorem 1 is thus completed.
\end{proof}
\section{Proof of Lemma 1 and Theorem 2}
\begin{proof}
    We first derive an upper-bound of the Euclidean distance between any two points in the simplex \(\Sigma_{i,t}^k\). Taking any \(\mathbf{q}_1\) and \(\mathbf{q}_2\in\Sigma_{i,t}^k\), the Euclidean distance between them is calculated as \(\left\|\mathbf{q}_1-\mathbf{q}_2\right\|_2\). By expanding the expression, we have
    \begin{equation}
        \left\|\mathbf{q}_1-\mathbf{q}_2\right\|_2=\sqrt{\left\|\mathbf{q}_1\right\|_2+\left\|\mathbf{q}_2\right\|_2-2\mathbf{q}_1^\top\mathbf{q}_2}.
    \end{equation}
    By the property that the \(l_2\) norm of a vector is upper-bounded by its \(l_1\) norm, we have
    \begin{equation}
        \left\|\mathbf{q}_1\right\|_2\leq\left\|\mathbf{q}_1\right\|_1=1.
    \end{equation}
    The equality follows from the property of the probability simplex. Furthermore, since the simplex only contains vectors with non-negative elements, we have \(\mathbf{q}_1^\top\mathbf{q}_2\geq0\) and thus \(-2\mathbf{q}_1^\top\mathbf{q}_2\leq0\). Combine the results above, we have
    \begin{equation}
        \left\|\mathbf{q}_1-\mathbf{q}_2\right\|_2=\sqrt{\left\|\mathbf{q}_1\right\|_2+\left\|\mathbf{q}_2\right\|_2-2\mathbf{q}_1^\top\mathbf{q}_2}\leq\sqrt{2}.
    \end{equation}
    Therefore, the Euclidean distance between any two points in this simplex is upper-bounded by \(\sqrt{2}\).\par
    Then, we prove the achievability of this upper-bound. Take any two vertices of this simplex, e.g., \((1,0,0,\cdots,0)\) and \((0,1,0,\cdots,0)\), where the corresponding  Euclidean distance is \(\sqrt{2}\). Therefore, the maximum Euclidean distance between any two points in the probability simplex \(\Sigma_{i,t}^k\) is \(\sqrt{2}\) for any \(i,t,k\).\par
    Subsequently, we analyze the maximum magnitude of the signal \(\hat{\mathbf{q}}_{i,t}^k\) for each class \(k\), i.e.,
    \begin{equation}
        \begin{aligned}
            \left\|\hat{\mathbf{q}}_{i,t}^k\right\|_2&=\left\|h_{i,t}P^k_{1,i,t}\sqrt{K}\mathbf{q}^k_{i,t}\right\|_2\leq\sqrt{K}\left|h_{i,t}P_{1,i,t}^k\right|\times\left\|\mathbf{q}^k_{i,t}\right\|_2\\
            &\leq\sqrt{K}\left|h_{i,t}P_{1,i,t}^k\right|\times\left\|\mathbf{q}^k_{i,t}\right\|_1=\sqrt{K}\left|h_{i,t}P_{1,i,t}^k\right|.
        \end{aligned}
    \end{equation}
    Accordingly, given that the maximum Euclidean distance between any two points in the probability simplex \(\Sigma_{i,t}^k\) is \(\sqrt{2}\), we can obtain the upper-bound of the \(l\)-2 sensitivity in (15), i.e.,
    \begin{equation}
    \begin{aligned}
        \Delta^k_{i,t}&=\max_{\mathcal{B}_i,\mathcal{B}'_i}\left\|\hat{\mathbf{q}}^k_{i,t}\left(\mathcal{B}_i\right)-\hat{\mathbf{q}}^k_{i,t}\left(\mathcal{B}'_i\right)\right\|_2\\
        &=\frac{1}{B_i}\left(\sum_{b\neq b'}\left\|\hat{\mathbf{q}}^k_{i,t}\left(\mathbf{u}_i^b\right)-\hat{\mathbf{q}}^k_{i,t}\left(\mathbf{u}_i^b\right)\right\|_2\right.\\
        &\quad\left.+\max_{\mathbf{u}_i^{b'},\hat{\mathbf{u}}_i^{b'}}\left\|\hat{\mathbf{q}}^k_{i,t}\left(\mathbf{u}_i^{b'}\right)-\hat{\mathbf{q}}^k_{i,t}\left(\hat{\mathbf{u}}_i^{b'}\right)\right\|_2\right)\\
        &\leq\frac{\sqrt{2K}\left|h_{i,t}P_{1,i,t}^k\right|}{B_i},
    \end{aligned}
    \end{equation}
    where \(\mathbf{u}_i^{b'}\) and \(\hat{\mathbf{u}}_i^{b'}\) are the specific altered sample in the neighboring datasets. The proof of Lemma 1 is thus completed.\par
    According to Theorem 1 in \cite{User}, to achieve \((\varepsilon_i,\delta_i)\)-DP of WD \(i\) after \(T\) training rounds, the standard deviation \(\sigma_i^k\) of the Gaussian DP noise imposed on the transmitted soft prediction for class \(k\) of WD \(i\) per round satisfies
    \begin{equation}
        \sigma_i^k=\frac{\Delta_{i,t}^k\sqrt{2T\ln\frac{1}{\delta_i}}}{\varepsilon_i}.
    \end{equation}
    Plugging in the obtained upper-bound of the sensitivity \(\Delta^k_{i,t}\) in (38), we have
    \begin{equation}
\sum_{j=1}^M\left|h_{j,t}P^k_{2,j,t}\right|^2+\sigma_n^2\geq\max_{i\in\mathcal{M}}4TK\left|h_{i,t}P_{1,i,t}^k\right|^2\rho_i,\forall k,t,
\end{equation}
where \(\rho_i\) is defined in Theorem 2. The proof of Theorem 2 is thus completed.
\end{proof}
\section{Proof of Proposition 1}
\begin{proof}
For each training round \(t\), we see that
\begin{equation}
\begin{aligned}
    &\sum_{i=1}^M\Omega_i\left(T,
    \left\{\mathcal{P}_t\right\}_{t=0}^{T-1}\right)=\sum_{i=1}^M\sum_{t=0}^{T-1}A_2\frac{\Phi_{1,i,t}^2+\Phi_{2,i,t}}{T^{\frac{3}{2}}}+\sum_{i=1}^M\sum_{t=0}^{T-1}\\
    &A_1\frac{\left(L_1\eta_t+1\right)\left\|\nabla F_i(\bm{\theta}_{i,t})\right\|_2\Phi_{1,i,t}}{\eta_tT^{\frac{3}{2}}}\stackrel{(a)}{\geq}\sum_{i=1}^M\sum_{t=0}^{T-1}A_2\frac{\Phi_{2,i,t}}{T^{\frac{3}{2}}},
\end{aligned}
\end{equation}
where \(A_1=6\gamma\eta_0L_2\) and \(A_2=6\eta_0\gamma^2L_2^2L_1\) are constant terms. The equality \((a)\) holds if \(\Phi_{1,i,t}=0,\forall i,t\), which leads to
\begin{equation}
    \frac{h_{i,t}P_{1,i,t}^k\sqrt{K}}{\lambda_t^k}-\frac{B_i^k}{B^k}=0,\quad\forall i,t,k.
\end{equation}
The optimal solution of \(P_{1,i,t}^k\) can be solved as
\begin{equation}
    P_{1,i,t}^{k^\ast} = \frac{B^k_i\lambda_t^k\overline{h}_{i,t}}{B^k\sqrt{K}\left|h_{i,t}\right|^2},\quad\forall i,t,k.
\end{equation}
Then, according to the peak transmit power constraint in (9), we have
\begin{equation}
\begin{aligned}
    \left(\lambda_t^k\right)^2&=\frac{\left(B^k\right)^2K\left|h_{i,t}P_{1,i,t}^k\right|^2}{\left(B_i^k\right)^2}\\
    &\leq\frac{\left(B^k\right)^2K\left|h_{i,t}\right|^2\left(P_i-\left|P_{2,i,t}^k\right|^2\right)}{\left(B_i^k\right)^2},\quad\forall i,t,k.
\end{aligned}
\end{equation}
With \(\left\{P^{k^\ast}_{1,i,t}, \forall i,k,t\right\}\) in (43), the optimization problem (P1) of the manuscript over \(\left\{\left\{P_{2,i,t}^k,\forall i\right\},\lambda_t^k, \forall t,k\right\}\) is given by
\begin{equation}
    \begin{aligned}
        \mbox{(P2)}\quad& \underset{\left\{P_{2,i,t}^k,\lambda_t^k,\forall i,t,k\right\}}{\text{min}}
& & \sum_{i=1}^M\sum_{t=0}^{T-1}A_2\frac{\Phi_{2,i,t}}{T^{\frac{3}{2}}}\\
 & \text{s.t.} & & \lambda_t^k\leq\Psi_{1,t}^k\left(B^k,\left\{P_i,h_{i,t},P_{2,i,t}^k,B^k_i,\forall i\right\}\right),\\
 &&&\quad \forall t,k,\\
 &&& \lambda_t^k\leq\Psi_{2,t}^k\left(B^k,\left\{h_{i,t},P_{2,i,t}^k,B^k_i,\rho_i,\forall i\right\}\right),\\
 &&&\quad\forall t,k.
    \end{aligned}
\end{equation}
Here
\begin{equation}
    \begin{aligned}
        \Psi_{1,t}^k\left(\cdot\right)&=\min_{i\in\mathcal{M}}\frac{B^k\left|h_{i,t}\right|\sqrt{K\left(P_i-\left|P_{2,i,t}^k\right|^2\right)}}{B^k_i},\\
        \Psi_{2,t}^k\left(\cdot\right)&=B^k\sqrt{\frac{\sum_{j=1}^M\left|h_{j,t}P^k_{2,j,t}\right|^2+\sigma_n^2}{4T\max_{i\in\mathcal{M}}\left(B_i^k\right)^2\rho_i}}.\nonumber
    \end{aligned}
\end{equation}
In the following, we solve the above problem in two cases.
\begin{itemize}
\item Case I: if \(\sigma_n^2\geq 4TK\left(\min_{i\in\mathcal{M}}|h_{i,t}|^2P_i\right)\max_{i\in\mathcal{M}}\rho_i\), the optimization problem (P2) becomes
\begin{equation}
    \begin{aligned}
        \mbox{(P2.1)}\quad& \underset{\left\{P_{2,i,t}^k,\lambda_t^k,\forall i,t,k\right\}}{\text{min}}
& & \sum_{i=1}^M\sum_{t=0}^{T-1}A_2\frac{\Phi_{2,i,t}}{T^{\frac{3}{2}}}\\
 & \text{s.t.} & & \lambda_t^k\leq\Psi_{1,t}^k\left(\cdot\right),\quad\forall t,k.
    \end{aligned}
\end{equation}
By analyzing the objective function of (46), we see that
\begin{equation}
    \begin{aligned}
        \sum_{i=1}^M\sum_{t=0}^{T-1}A_2\frac{\Phi_{2,i,t}}{T^{\frac{3}{2}}}&=\sum_{i=1}^M\sum_{t=0}^{T-1}\frac{A_2\sum_{k=1}^K\frac{B^k_i}{B_i}\mathbb{E}\left[\left\|\hat{\mathbf{m}}_t^k\right\|^2_2\right]}{\left(\lambda_t^k\right)^2T^{\frac{3}{2}}}\\
        &=\sum_{i=1}^M\sum_{t=0}^{T-1}A_2\sum_{k=1}^K\frac{B^k_i}{B_i}K\left(\frac{\sigma_n^2}{\left(\lambda_t^k\right)^2T^{\frac{3}{2}}}\right.\\
        &\quad\left.+\frac{\sum_{j=1}^M\left|h_{j,t}P^k_{2,j,t}\right|^2}{\left(\lambda_t^k\right)^2T^{\frac{3}{2}}}\right).
    \end{aligned}
\end{equation}
The objective function in (47) is monotonically increasing with respect to \(P_{2,i,t}^k,\forall i,t,k\). Therefore, the optimal transmit power associated with artificial noises is \(P^{k^\ast}_{2,i,t}=0,\forall i,t,k\). With the optimal \(\left\{P_{2,i,t}^k,\forall i,t,k\right\}\), the objective function in (P2.1) is monotonically decreasing with respect to \(\lambda_t^k,\forall t,k\), which yields
\begin{equation}
\lambda_t^{k^\ast}=\min_{i\in\mathcal{M}}\frac{B^k\sqrt{K}\left|h_{i,t}\right|\sqrt{P_i}}{B_i^k},\quad\forall t,k.
\end{equation}
\item Case II: if \(\sigma_n^2< 4TK\left(\min_{i\in\mathcal{M}}|h_{i,t}|^2P_i\right)\max_{i\in\mathcal{M}}\rho_i\), the optimization problem (P2) becomes
\begin{equation}
    \begin{aligned}
        \mbox{(P2.2)}\quad& \underset{\left\{P_{2,i,t}^k,\lambda_t^k,\forall i,t,k\right\}}{\text{min}}
& & \sum_{i=1}^M\sum_{t=0}^{T-1}A_2\frac{\Phi_{2,i,t}}{T^{\frac{3}{2}}}\\
 & \text{s.t.} && \lambda_t^k\leq\Psi_{2,t}^k\left(\cdot\right),\quad\forall t,k.\nonumber
    \end{aligned}
\end{equation}
Notice that
\begin{equation}
\frac{\sum_{j=1}^M\left|h_{j,t}P^k_{2,j,t}\right|^2+\sigma_n^2}{\left(\lambda_t^k\right)^2T^{\frac{3}{2}}}\stackrel{(b)}{\geq}\frac{4T\left(\lambda^k_t\right)^2\max_{i\in\mathcal{M}}\left(B_i^k\right)^2\rho_i}{\left(B^k\lambda^k_t\right)^2T^{\frac{3}{2}}},\nonumber
\end{equation}
To achieve the equality in \((b)\), the optimal \(\left\{\left\{P_{2,i,t}^k,\forall i\right\},\lambda_t^k, \forall t,k\right\}\) satisfies the following equation,
\begin{equation}
\sum_{j=1}^M\left|h_{j,t}P^{k^\ast}_{2,j,t}\right|^2+\sigma_n^2=4T\left(\lambda_t^{k^\ast}\right)^2\max_{i\in\mathcal{M}}\left(\frac{B_i^k}{B^k}\right)^2\rho_i,\;\forall t,k.\nonumber
\end{equation}
\end{itemize}
The proof of Proposition 1 is thus completed.
\end{proof}
\section{Proof of Proposition 2}
\begin{proof}
Given the optimal transceiver design per round, we analyze the long-term optimization problem for the training rounds decision \(T\). According to the optimal transceiver design in Proposition 1, the \(i\)-th argument of the objective function \(\Omega_i\left(T\right)\) with respect to \(T\) in Problem (P1) of the manuscript is given by
\begin{equation}
    \Omega_i\left(T\right)=\begin{cases}
        \frac{3f_{i,max}}{\eta_0\sqrt{T}}+A_2\sum_{t=0}^{T-1}&\sum_{k=1}^K\frac{K\sigma_n^2B^k_i}{B_i\left(\lambda_t^{k^\ast}\right)^2T^{\frac{3}{2}}},\\
        &\text{if }\quad T\leq T_0,\\
        \frac{3f_{i,max}}{\eta_0\sqrt{T}}+A_2\sum_{k=1}^K&\frac{4\sqrt{T}KB^k_i\max_{i\in\mathcal{M}}\left(B_i^k\right)^2\rho_i}{B_i\left(B^k\right)^2},\\
        &\text{if}\quad T> T_0,\nonumber
    \end{cases}
\end{equation}
where \(T_0=\frac{\sigma_n^2}{4K\left(\min_{i\in\mathcal{M}}|h_{i,t}|^2P_i\right)\max_{i\in\mathcal{M}}\rho_i}\). To find the optimal number of training rounds \(T^\ast\), we first discuss the optimization problems for different regions of \(T\).
\begin{itemize}
    \item Case I: \(T\leq T_0\)\par
    The optimization problem in this case is given by
    \begin{equation}
    \begin{aligned}
        \mbox{(P3.1)}\quad&\underset{T}{\text{min}}&&
\sum_{i=1}^M\left(\frac{3f_{i,max}}{\eta_0\sqrt{T}}+A_2\sum_{t=0}^{T-1}\sum_{k=1}^K\frac{K\sigma_n^2B^k_i}{B_i\left(\lambda_t^{k^\ast}\right)^2T^{\frac{3}{2}}}\right)\\
 & \text{s.t.} & &T\leq T_0.\nonumber
    \end{aligned}
    \end{equation}
    According to (48), we have
    \begin{equation}
        \begin{aligned}
            \sigma_n^2&\geq 4TK\left(\min_{i\in\mathcal{M}}\frac{|h_{i,t}|^2P_i}{\left(B_i^k\right)^2}\right)\left(\max_{i\in\mathcal{M}}\left(B_i^k\right)^2\rho_i\right)\\
            &=4T\left(\lambda_t^{k^\ast}\right)^2\max_{i\in\mathcal{M}}\left(\frac{B_i^k}{B^k}\right)^2\rho_i.
        \end{aligned}
    \end{equation}
    By (49), we can see the objective function in (P3.1) is lower-bounded, i.e.,
    \begin{equation}
    \begin{aligned}
        \sum_{i=1}^M\Omega_i\left(T\right)&\geq\sum_{i=1}^M\left(\frac{3f_{i,max}}{\eta_0\sqrt{T}}\right.\\
        &\quad\left.+A_2\sum_{k=1}^K\frac{4\sqrt{T}KB^k_i\max_{i\in\mathcal{M}}\left(B_i^k\right)^2\rho_i}{B_i\left(B^k\right)^2}\right).
    \end{aligned}
    \end{equation}
    The lower-bound in (50) is exactly the expression of the objective function when \(T>T_0\). Therefore, the objective value associated with the optimal \(T\) obtained by solving (P3.1) is not smaller than that obtained by minimizing the objective function when \(T>T_0\).
    \item Case II: \(T> T_0\)\par
    The optimization problem in this case is given by
    \begin{equation}
    \begin{aligned}
        \mbox{(P3.2)}\quad&\underset{T}{\text{min}}&&
\frac{3\sum_{i=1}^Mf_{i,max}}{\eta_0\sqrt{T}}\\
&&&\quad+\sqrt{T}A_2\sum_{i=1}^M\sum_{k=1}^K\frac{4KB^k_i\max_{i\in\mathcal{M}}\left(B_i^k\right)^2\rho_i}{B_i\left(B^k\right)^2}\\
&\text{s.t.}&& T>T_0.
\nonumber
    \end{aligned}
    \end{equation}
    We can see that the objective function of (P3.2) is in the form of \(\frac{a}{\sqrt{T}}+b\sqrt{T}\) with \(a,b>0\). It is a convex function and the closed-form optimal solution of the non-negative and continuous \(\hat{T}\) is given by
    \begin{equation}
    \begin{aligned}
        \hat{T}^\ast&=\frac{a}{b}=\frac{3\sum_{i=1}^Mf_{i,max}}{\eta_0A_2\sum_{i=1}^M\sum_{k=1}^K\frac{4KB^k_i\max_{i\in\mathcal{M}}\left(B_i^k\right)^2\rho_i}{B_i\left(B^k\right)^2}}\\
        &=\frac{3\sum_{i=1}^Mf_{i,max}}{4\eta_0A_2M\max_{i\in\mathcal{M}}\rho_i\sum_{k=1}^K\left(\frac{B_i^k}{B^k}\right)^2}.
    \end{aligned}
    \end{equation}
    Then, we round the continuous \(\hat{T}^\ast\) to the nearest integer and the proof of Proposition 2 is thus completed.
\end{itemize}
\end{proof}
}


\begin{thebibliography}{99}
\bibliographystyle{IEEEtran}
\bibitem{FL}
B. McMahan, E. Moore, D. Ramage, S. Hampson, and B. A. Y. Arcas, ``Communication-efficient learning of deep networks from decentralized data,” in {\it{Proc. 20th AISTATS,}} vol. 54, Apr. 2017, pp.1273–1282.
\bibitem{OTA}
G. Zhu, Y. Wang, and K. Huang, ``Broadband Analog Aggregation for Low-Latency Federated Edge Learning," in {\it{IEEE TWC,}} vol. 19, no. 1, pp. 491-506, Jan. 2020.
\bibitem{DP}
C. Dwork, and A. Roth, {\it{The Algorithmic Foundations of Differential Privacy,}} now, 2014.
\bibitem{DP-FL}
M. S. E. Mohamed, W.-T. Chang, and R. Tandon, ``Privacy Amplification for Federated Learning via User Sampling and Wireless Aggregation," in {\it{IEEE JSAC,}} vol. 39, no. 12, pp. 3821-3835, Dec. 2021.
\bibitem{D-FL}
Z. Hu, J. Yan, and Y.-J. A. Zhang, ``Communication-Learning Co-Design for Differentially Private Over-the-Air Federated Learning with Device Sampling," in {\it{IEEE TWC,}} vol. 23, no. 11, pp. 16788-16804, Nov. 2024.
\bibitem{OTA-FD}
Z. Hu, J. Yan, Y.-J. A. Zhang, J. Zhang, and K. B. Letaief, ``Optimal Transceiver Design in Over-the-Air Federated Distillation," {\it{arXiv preprint arXiv:2507.15256,}} 2025.
\bibitem{S-FD}
J. Shao, F. Wu, and J. Zhang, “Selective knowledge sharing for privacy-preserving federated distillation without a good teacher,” {\it{Nature Commun.,}} vol. 15, no. 1, pp. 1–11, Jan. 2024.
\bibitem{User}
K. Wei, {\it{et al.}}, “User-Level Privacy-Preserving Federated Learning: Analysis and Performance Optimization," in {\it{IEEE TMC,}} vol. 21, no. 9, pp. 3388-3401, 1 Sept. 2022.
\bibitem{CSI}
L. Zhao, H. Xu, Z. Wang, X. Chen, and A. Zhou, “Joint Channel Estimation and Feedback for mm-Wave System Using Federated Learning,” in {\it{IEEE Commun. Letters,}} vol. 26, no. 8, pp. 1819-1823, Aug. 2022.
\bibitem{MA}
A. Martín, A. Chu, I. Goodfellow, H. B. McMahan, I. Mironov, K. Talwar, and L. Zhang, “Deep learning with differential privacy,” in {\it{Proc. ACM CCS,}} Vienna, Austria, Oct. 2016, pp. 308–318.
\bibitem{OFDM}
A. Haque, P. Kumar, and A. K. Singh, ``IEEE 802.11ac: 5th generation wifi networking,” in {\it{WAP,}} vol. 2, no. 4, pp. 235–241, 2012.
\bibitem{ref12}
X. Wang, S. Magnússon, and M. Johansson, ``On the convergence of step decay step-size for stochastic optimization," {\it{Advances in Neural Information Processing Systems,}} vol. 34, pp. 14226-14238, 2021.
\end{thebibliography}
\end{document}